\documentclass[10pt,aps,prc,twocolumn,showpacs,showkeys,nofootinbib,superscriptaddress]{revtex4-2}
\usepackage{graphicx}
\usepackage{amsmath,amsfonts,mathrsfs}
\usepackage{enumitem} 

\graphicspath{{./img/}}
\def\vec#1{\boldsymbol{ #1}}

\def\Tu{T_{\rm U}}

\def\tcoll{t_{coll}}
\def\tdiff{t - \tcoll}
\def\Tdiff{T^2-\Tu^2}

\newcommand{\jinr}{Joint Institute for Nuclear Research, Joliot-Curie str. 6, Dubna 141980, Russia}
\newcommand{\kurchatov}{NRC Kurchatov Institute, Moscow, Russia}

\usepackage[colorlinks = true,
linkcolor = blue,
urlcolor  = blue,
citecolor = blue,
anchorcolor = blue]{hyperref}

\usepackage{slashed}

\begin{document}
	
\title{Modeling of acceleration in heavy-ion collisions: occurrence of temperature below the Unruh temperature}

\author{G. Yu. Prokhorov}
\email{prokhorov@theor.jinr.ru}
\affiliation{\jinr}
\affiliation{\kurchatov}
\author{D. A. Shohonov}
\email{d.shohonov@gmail.com}
\affiliation{\jinr}
\author{O. V. Teryaev}
\email{teryaev@jinr.ru}
\affiliation{\jinr}
\affiliation{\kurchatov}
\author{N. S. Tsegelnik}
\email{tsegelnik@jinr.ru}
\affiliation{\jinr}
\author{V. I. Zakharov}
\email{vzakharov@itep.ru}
\affiliation{\kurchatov}
\affiliation{\jinr}

\begin{abstract}
It has recently been shown that extremely strong electric fields can be created in central collisions of heavy ions, due to which the Schwinger effect can be significant. A direct analogue of the electric field in hydrodynamics is the acceleration of the medium. Using the parton-hadron-string dynamics (PHSD) framework we model the Au-Au collisions at intermediate collision energies $\sqrt{s_{NN}}=4.5-11.5\,$GeV and obtain the spatial distribution of acceleration at different time moments. The present study demonstrates that extremely high acceleration of the order of 1 GeV may be generated in both central and non-central collisions, and that the distribution exhibits a core-corona structure. Consequently, in contrast to the case with an electric field, the Unruh effect is expected to be significant. 
It is demonstrated that for the confined phase the temperature is less than the Unruh temperature.
Conversely, in the deconfined phase, the relationship is inverse. The obtained results thus support the prediction about the existence of states with $T<T_U$ at the early stages of the collision and the associated complementary description of thermalization in terms of the novel phase transition at the Unruh temperature.
\end{abstract}

\maketitle


\section{\label{sec:level1}Introduction}

In recent years the field of research of relativistic heavy ion collisions has undergone rapid development, initiated by the active operation of the Relativistic Heavy Ion Collider (RHIC) and, later, of a related program at the Large Hadron Collider (LHC). The discovery of a new state of nuclear matter~ \cite{BRAHMS:2004adc, PHOBOS:2004zne, STAR:2005gfr, PHENIX:2004vcz, ALICE:2022wpn}, known as the quark-gluon plasma (QGP), has prompted a rethinking of the initial assumptions regarding the behavior of high-temperature quantum chromodynamics (QCD) matter. Specifically, the prevailing expectations of a weakly coupled gas-like state~\cite{PhysRevLett.30.1346,PhysRevD.8.3633,PhysRevD.9.980,Shuryak:1988ck} have been challenged, suggesting instead a behavior more akin to a strongly-coupled almost ideal fluid~\cite{Karsch:2000kv, PhysRevLett.87.081601, MOLNAR2002495, DATTA2003487, PhysRevLett.92.012001}.

It is believed that the QGP is formed at the earliest stage of the system evolution in the collisions of relativistic nuclei, followed by the process of hadronization~\cite{Florkowski:2014yza, Satz:2018oiz}. Such collisions have been shown to generate substantial electromagnetic fields~\cite{Skokov:2009qp, Bali:2011qj, PhysRevC.83.054911, Deng:2012pc, Miransky:2015ava,ToneevRogachevskyVoronyuk2016}. Strong magnetic fields may lead to the so-called Chiral magnetic effect~\cite{FukushimaKharzeevWarringa2008, SadofyevShevchenkoZakharov2011, ToneevKonchakovskiVoronyukBratkovskayaCassing2012, Kharzeev:2013ffa, Kharzeev:2024zzm}, which is of interest from both a theoretical perspective, primarily due to its deep relationship with quantum chiral anomaly~\cite{FukushimaKharzeevWarringa2008,SadofyevShevchenkoZakharov2011,Landsteiner:2016led}, and from an experimental perspective~\cite{Li:2014bha,Zhao:2019hta, Li:2020dwr, Ong:2020ffe}. Recent work~\cite{TayaNishimuraOhnishi2024} has shown that electric fields of the order of the QCD scale arise in central nuclear collisions, suggesting that it is possible to study the Schwinger effect directly in such strong electric fields.

Early experiments~\cite{PhysRevLett.47.229, Anikina:1984, STAR:2007ccu} initiated an active theoretical study~\cite{Ayala:2001jp, Liang:2004ph, Liang:2004xn, Gao:2007bc, Becattini:2013fla, Pang:2016igs, Karpenko:2016jyx, Sorin:2016smp, Li:2017slc} of the polarization phenomena in the heavy ion collisions. Recent experimental observations~\cite{Adamczyk2017, STAR:2018gyt, PhysRevC.101.044611, 2022137506} have demonstrated the non-zero global polarization of hyperons with a tendency to decrease with increasing collision energy. The underlying mechanism is usually related to the hydrodynamic vorticity~\cite{Adamczyk2017, Becattini:2020ngo}, the structure of which has been extensively studied in~\cite{PhysRevC.93.031902, PhysRevC.93.064907, Jiang:2016woz, Karpenko:2016jyx, Xia:2018tes, Ivanov:2018eej, Kolomeitsev:2018svb, Tsegelnik:2022eoz}. It is worth to be noted that the maximum values of global polarization and averaged vorticity are predicted to occur at intermediate energies, where the forthcoming Nuclotron-based Ion Collider fAcility (NICA) experiment~\cite{Drnoyan:2020acj, Nazarova:2024jic} will be particularly well-suited. The rotation of the nuclear matter is also investigated within the framework of lattice QCD~\cite{Yamamoto:2013zwa, Braguta:2020biu, Braguta:2021jgn}, effective models~\cite{Jiang:2016wvv, Chernodub:2016kxh, Wang:2018sur, Fujimoto:2021xix} and holographic approaches~\cite{Arefeva:2020jvo, Chen:2020ath, Braga:2022yfe, GolubtsovaTsegelnik2023}.

From the point of view of hydrodynamics, the direct analogue of the magnetic field $B_{\mu} = \frac{1}{2} \varepsilon_{\mu\nu\alpha\beta} u^{\nu} F^{\alpha\beta}$ is vorticity $\omega_{\mu} = \frac{1}{4} \varepsilon_{\mu\nu\alpha\beta} u^{\nu} \Omega^{\alpha\beta}$, and the direct analogue of the electric field $E_{\mu} = F_{\mu\nu} u^{\nu}$ is acceleration $a_{\mu} = - \Omega_{\mu\nu} u^{\nu}$ (here $F_{\mu\nu} = \partial_{\mu} A_{\nu} - \partial_{\nu} A_{\mu}$ is the usual electromagnetic field tensor, and $\Omega_{\mu\nu} = \partial_{\mu} u_{\nu} - \partial_{\nu} u_{\mu}$ is the kinematic vorticity tensor, while $u^{\mu}$ is the four-velocity of the medium), being, respectively, the vector ($E_{\mu}$ and $a_{\mu}$) and pseudovector ($B_{\mu}$ and $\omega_{\mu}$) components of corresponding tensors ($F_{\mu\nu}$ and $\Omega_{\mu\nu}$). And while vorticity and electric fields have been studied in details, there are not many works devoted to modeling acceleration (see, for example \cite{PalermoGrossiKarpenkoBecattini2024, Karpenko:2018erl}). Of particular interest is the study of acceleration effects using a first-principle lattice approach, since in this case the well-known sign problem can be avoided; the first steps were recently made in this direction \cite{Ambrus:2023smm}.

The present research aims to examine the process of generation and evolution of acceleration in both central and non-central heavy-ion collisions simulated within the PHSD model. The focus is on modeling spatial distributions of acceleration at different time moments of the system evolution, with the intention of evaluating acceleration (and the corresponding Unruh temperature) and comparing it with the temperature from the equation of state (EoS). The existence of extremely high accelerated zones, consisting of hadrons, where the Unruh temperature exceeds that obtained from the EoS, will be demonstrated. 
Furthermore, it will be shown that the partonic phase is characterized by minimal acceleration, which results in the temperature being above the Unruh temperature
\footnote{Let us emphasize that the difference between the temperature and the Unruh temperature is not at all connected with any ``violation'' of the Unruh effect, as it might naively seem. According to the Unruh effect, the temperature is equal to the Unruh temperature only for the concrete quantum (vacuum) state, as discussed below.}.

Let us recall that according to the Unruh effect~\cite{Unruh1976}, the Minkowski vacuum for an accelerated observer looks like a thermal medium (or ``bath'') with a so-called Unruh temperature
\begin{align}\label{eq:Tu}
	\Tu = \frac{a}{2\pi},
\end{align}
where $a=\sqrt{-a_{\mu}a^{\mu}}$ is the modulus of acceleration of the observer or detector \footnote{In our case, the medium plays the role of a ``detector'' and acceleration will be related to the medium.}.  The Unruh effect has already been discussed in high-energy collisions in the context of thermal hadron production. In particular, a series of papers~\cite{CleymansSatz1993, KharzeevTuchin2005, Kharzeev2006, CastorinaKharzeevSatz2007, BecattiniCastorinaManninenSatz2008, Becattini:2009sc, Castorina:2014fna}, proposed a mechanism for thermal hadron production based on the idea that the color confinement forms an event horizon for quarks and gluons, similar to the event horizon of a black hole, which can be crossed by quantum tunneling, providing a QCD analogue of Hawking radiation. The effective radiation temperature is determined by the chromodynamic force and is $T \approx Q_s/2\pi$, where $Q_s$ is the saturation momentum of gluons, which characterizes the strength of the color fields~\cite{Castorina:2014fna}.

Recently, this idea was developed in connection with a study of the phase diagram in the $(a,T)$ axes~\cite{Prokhorov:2023dfg} (see also~\cite{Chernodub:2024wis}). Let's take a brief detour and discuss the prehistory of the issue. In systems with finite proper temperature and acceleration, $T$ and $a$ can be considered as independent parameters. In this case, the point $T = \Tu$ corresponds to the state of Minkowski vacuum. The diagram $(a,T)$ was initially considered in~\cite{Becattini2018}, where it was suggested that the temperature $T = \Tu$ is the minimal one for an accelerated medium. This statement was developed in~\cite{ProkhorovTeryaevZakharov2019}, where it was shown that the properties of the medium with $T < \Tu$ change qualitatively. In~\cite{Prokhorov:2023dfg}, within the framework of a simple model of Dirac fields in the Euclidean Rindler space, it was shown that the temperature $T = \Tu$ is critical. The transition through this point is accompanied by a jump in heat capacity, which indicates a second-order phase transition. The states with $T<\Tu$ can be investigated through analytical continuation, and turn out to be very unusual. In the language of the Euclidean Rindler space, they correspond to a cone with an angle greater than $2\pi$ (i.e. negative angular deficit). 

Moreover, in~\cite{Prokhorov:2023dfg} it was suggested that such states should be formed during the early stages of heavy ion collisions, when the medium experiences a large deceleration, due to stopping forces, but has not yet had time to thermalize. Taking into account these states (which are unstable and decay) in collisions of heavy ions makes it possible to connect the mentioned phase transition with thermalization (for more details see Section~\ref{sec:interpretation}). 

Thus, there is a strong theoretical motivation for modeling acceleration in heavy ion collisions, and, as we will see, the mentioned prediction that states with $T<\Tu$ are realized at the initial time moments after collisions is confirmed.


The paper has the following structure. Section~\ref{sec:PHSD} describes how the PHSD model can be used to simulate acceleration and temperature in heavy ion collisions. Section~\ref{sec:result} presents the results of modeling. Section~\ref{sec:interpretation} is devoted to the theoretical interpretation of the results obtained in connection with the Unruh effect and the phase transition at the Unruh temperature. The conclusions are formulated in Section~\ref{sec:conclusions}.

\section{Calculating acceleration and temperature within PHSD}\label{sec:PHSD}

The PHSD model has been successfully applied to the quantitative analysis of the observables in heavy-ion collisions in the energy range from SIS to upper RHIC energies~\cite{CassingBratkovskaya2009, BratkovskayaCassingKonchakovskiLinnyk2011, Linnyk:2015rco}. Let us briefly describe how the model is used to calculate the properties of the medium formed in nuclear collisions. The PHSD framework generates distributions of test particles at each moment of time with momenta $p_{i_h}(t)$ and coordinates $r_{i_h}(t)$ (the indices $i$ and $h$ indicate the $i$-th particle of type $h$). In order to characterize the continuous medium, the stress-energy tensor is used, the expression of which is
\begin{align}\label{eq:Tmunu}
	T^{\mu\nu}(x) = \int \frac{d^3 p}{(2\pi)^3} \frac{p^\mu p^\nu}{p^0} f(x,p),
\end{align}
where $f(x, p)$ is the distribution function. This simple form of the tensor satisfies our problem, since all the dynamics of the system are hidden in the transport equations, and we are interested in the properties of the medium at specific moments of time. Formally, the distribution function for test particles of type $h$ may be written as
\begin{align}\label{eq:f-distr}
	f^{(h)}(x,p) &=
	\sum_{i_h} (2\pi)^3 \delta^{(3)}\big(\vec{p} - \vec{p}_{i_h}(t)\big)
	\nonumber\\
	&\quad\times
	\delta\big(p_0-\sqrt{m_h^2+\vec{p\,}^2}\big)\delta^{(3)}\big(\vec{r}-\vec{r}_{i_h}(t)\big),
\end{align}
where $\vec{r}$ and $\vec{p}$ are the coordinate and momentum in phase space, respectively.

The particles of colliding nuclei may be divided into spectators and participants, the former of which fly past each other without interacting during the collision of nuclei. In addition, all particles produced during the system evolution are considered to be participants. We assumed that only the participant particles would constitute the medium, i.e. they were taken into account when calculating the energy-momentum tensor~\eqref{eq:Tmunu}. It is important to note that, although the spectators are excluded from the filling of the tensor~\eqref{eq:Tmunu}, they actually influence the dynamics of the system at each moment of time, which is taken into account in the transport equations of the PHSD model.

Finally, to perform a transition from the distribution of test particles to a continuous medium, the so-called smearing function $\Phi(\vec{r}, \vec{r}_{i_h}(t))$ is introduced instead of the spatial $\delta$-function in~\eqref{eq:f-distr}:
\begin{align}
	f^{(h)}(x,p) &= \frac{1}{\mathcal{N}}
	\sum_{i_h} (2\pi)^3 \delta^{(3)}\big(\vec{p} - \vec{p}_{i_h}(t)\big)
	\nonumber\\
	&\quad\times
	\delta\big(p_0-\sqrt{m_h^2+\vec{p\,}^2}\big) \Phi(\vec{r}, \vec{r}_{i_h}(t)),
\end{align}
where $\mathcal{N} = \int d^3r \Phi(\vec{r}, \vec{r}_{i_h}(t))$ is the normalization factor.  Setting a spatial grid with steps $(\Delta  x, \Delta  y,\Delta z) = (1,1,1/\gamma)\,$fm, where $\gamma=1/\sqrt{1-v^2}$ is the Lorentz factor of the colliding ions, the energy-momentum tensor now looks as follows:
\begin{align}\label{eq:Tmunu-sum}
	T^{\mu\nu}(x) = \frac{1}{\mathcal{N}} \sum_{h, i_h} \frac{p^\mu_{i_h}(t) p^\nu_{i_h}(t)}{p^0_{i_h}(t)} \Phi(\vec{r}, \vec{r}_{i_h}(t)).
\end{align}

After making the transition to a continuous medium, we can define the flow velocity $u_\mu$ and the local energy density $\varepsilon$ in the Landau system as an eigenvector and the corresponding eigenvalue of the energy-momentum tensor~\eqref{eq:Tmunu-sum}:
\begin{align}\label{eq:Tmunu-eigen}
	u_\mu T^{\mu\nu} = \varepsilon u^\nu.
\end{align}
To determine the local temperature, we use the EoS from \cite{SatarovDmitrievMishustin2009} and solve $T(n_B, \varepsilon) = T$, where $\varepsilon$ is taken from~\eqref{eq:Tmunu-eigen} and $n_B$ is the baryon density, calculated from the baryon current $J^\mu_B$ as $n_B = u_\mu J^\mu_B$\footnote{Note that this equation should not be confused with the condition $J^\mu_B= n_B u_\mu $, that is, the current is diffusion-less, which fixes the Eckart frame.}.
Although this EoS is not the most modern, it has proven itself well and has been well tested, in particular, in calculating global polarization. In this paper, we use the hadronic EoS from \cite{SatarovDmitrievMishustin2009}, which does not contain a phase transition to the parton phase \footnote{It should also be noted that PHSD itself has its own built-in equation of state, so the parton phase is described correctly in the kinetic equations.
And the equation of state discussed here does not participate in the equations of motion in any way, and is only
needed to estimate the temperature.}. At the same time, the phase transition is crucial only for the areas of high density and temperature, observed in the first moments after the collision, where, as will be shown, the Unruh temperature will be less than the temperature from EoS (while we are primarily interested just in the areas where the Unruh temperature will be higher than the EoS temperature). And the most difference in the use of other EoS can be only in the area between the core and the corona. Moreover, generally speaking, the temperatures are not much higher than the critical one, which allows us to limit ourselves to this EoS at least at the first stage.

Details of the algorithm used, the explicit form of the smearing kernel $\Phi$, the spectator separation procedure, the analysis of the resulting medium, and the application to the calculation of the global polarization can be found in a series of papers~\cite{Tsegelnik:2022eoz, Voronyuk:2023vyu, TsegelnikKolomeitsevVVoronuykFreezeOut2023, Tsegelnik:2024ruh}.

Once we know the flow velocity $u_\mu$, the acceleration can be calculated using the formula
\begin{align}\label{eq:accel}
	a_\mu = u^\nu \partial_\nu u_\mu
\end{align}
at different times at the points of the spatial grid. Then, the Unruh temperature can be evaluated using~\eqref{eq:Tu}.

The PHSD transport model employs the parallel ensemble method, which involves the simultaneous simulation processing of $N$ collisions. To improve the accuracy of the calculated values, the PHSD model is initialized $M$ times, with the fluidization procedure being performed for each run individually. Consequently, the total number of simulated events is $N \times M$. In this paper, the number of parallel ensembles was set to $N = 200$ for the collision energies of $7.7$ and $11.5\,$GeV, and the number of independent model runs is $M \approx 500$. Thus, the total statistics is $(5 - 10) \times 10^4$ events for each impact parameter and collision energy.

\section{Results}\label{sec:result}

In this sections, we present a detailed analysis of temperature $T$ (obtained from the EoS), the Unruh temperature $\Tu$ (calculated as \eqref{eq:Tu} and \eqref{eq:accel}), and the difference $\Tdiff$ for the medium produced in the central ($b = 0\,$fm) and off-center ($b = 7.5\,$fm) heavy-ion collisions.

\subsection{Temperature and acceleration in central and off-center collisions: overall analysis}

\begin{figure*}
	\includegraphics[width=0.48\linewidth]{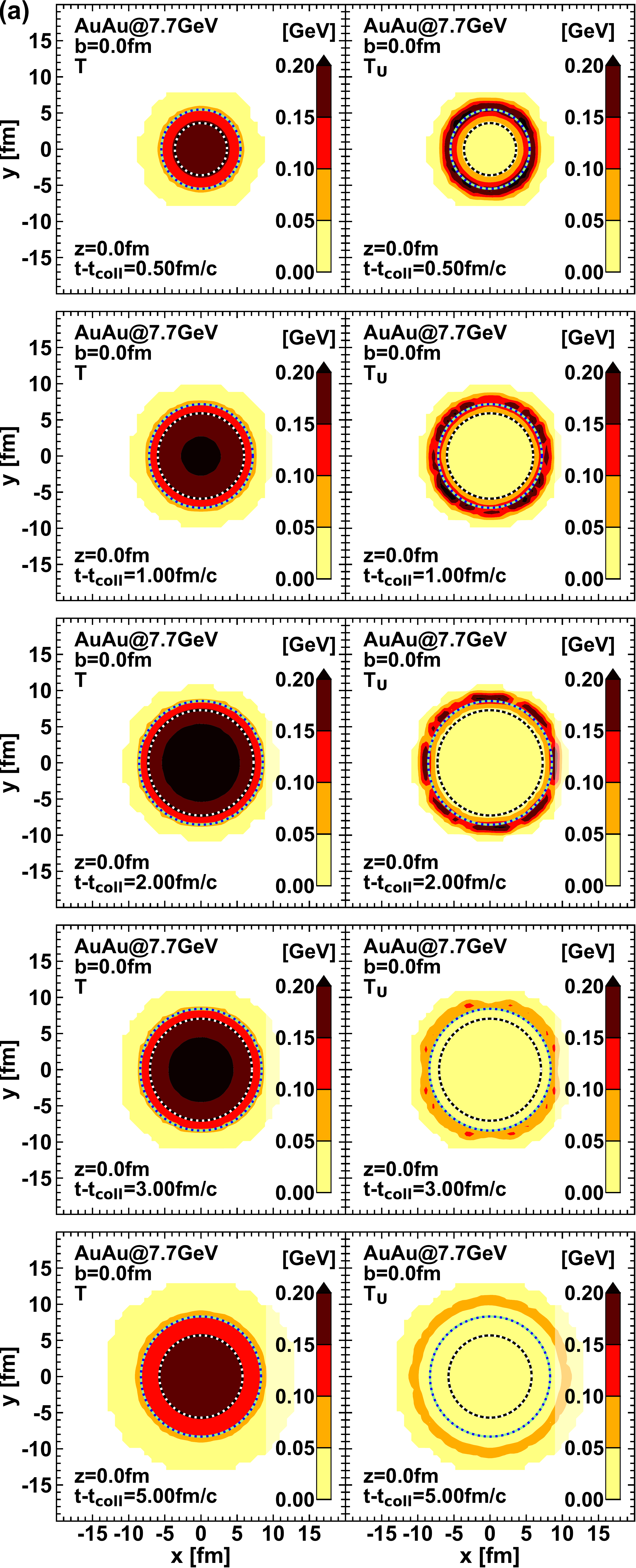}\hfill
	\includegraphics[width=0.48\linewidth]{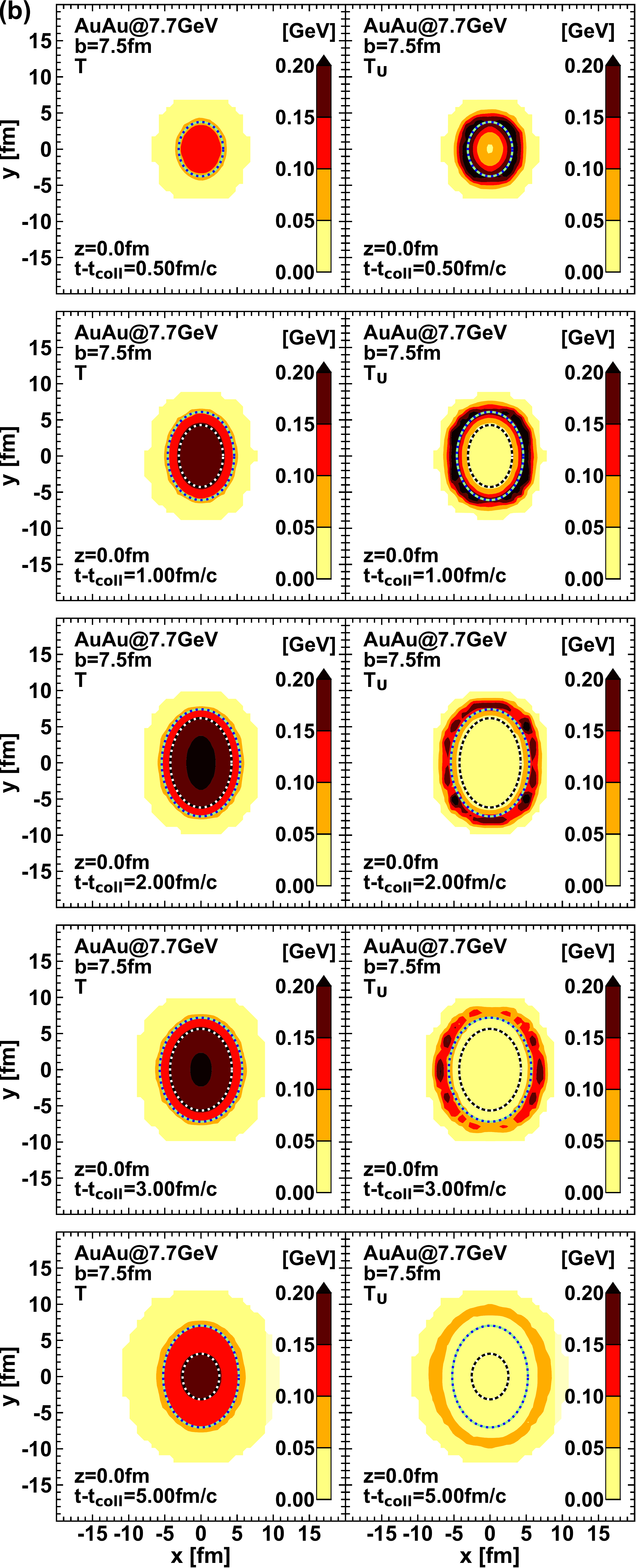}
	\caption{Profiles of temperature $T$ (first column), characteristic Unruh temperature $\Tu$ (second column) in the $z = 0$ plane, calculated at different times after the collision for Au + Au collisions at $\sqrt{s_{NN}} = 7.7\,$GeV. The profiles of $T$ and $\Tu$ shown in (a) are calculated for a central collision ($b = 0\,$fm); in (b) for an off-center collision ($b = 7.5\,$fm). The dotted curves show two contours in the $xy$ plane, limiting the medium in energy density to the value $\varepsilon = 500\,$MeV/fm$^3$ (inner contour) and $\varepsilon = 50\,$MeV/fm$^3$ (outer contour).}
	\label{fig:T-and-Tu}
\end{figure*}

Figure~\ref{fig:T-and-Tu} shows the spatial profiles of the temperature $T$ and the Unruh temperature $\Tu$ in the $z = 0$ plane, calculated for Au + Au collisions at the energy $\sqrt{s_{NN}} = 7.7\,$ GeV for various times.
The time $\tdiff = 0\,$fm$/c$ corresponds to the moment of contact of the nuclei. At this moment, a hot medium begins to form in the center of the system, which then expands. The volume occupied by the hot substance changes non-trivially~\cite{TsegelnikKolomeitsevVVoronuykFreezeOut2023}, and the temperature of the resulting medium reaches its highest value at the moment of maximum overlap of the nuclei ($\tdiff \approx 2.5\,$fm$/c$). The temperature profiles $T$ in Figure~\ref{fig:T-and-Tu} show that the hottest medium in the central cut is formed in the time interval from 2 to 3 fm$/c$ after the moment of contact of the nuclei for the collision under consideration. After 3 fm$/c$, the resulting medium cools down due to hadronization and thermalization processes~\cite{Florkowski:2014yza,CastorinaKharzeevSatz2007,BecattiniCastorinaManninenSatz2008,Becattini:2009sc}. As can be seen from the temperature profiles $T$, the hot medium in the plane $z = 0\,$fm is located inside the contour that limits the medium in energy density to the value $\varepsilon = 50\,$MeV/fm$^3$. In what follows, by a fireball we mean a medium limited by this energy density ($\varepsilon \geq 50\,$MeV/fm$^3$). In the PHSD model, the QGP is defined to exist inside a region limited by the energy density to a value of $\varepsilon = 500\,$MeV/fm$^3$ from below (see~\cite{CassingBratkovskaya2009} and ref. therein). In Figure~\ref{fig:T-and-Tu} we denoted this region by dashed black and white contour. Note that such an analysis is valid for a short period of time after the collision ($\tdiff = 3 - 4\,$fm$/c$). In the PHSD model, after the maximum overlap of nuclei the fraction of the partonic phase rapidly decreases and becomes insignificant~\cite{MoreauSolovevaGrishmanovskiiVoronyukOlivaSongKireyeuCociBratkovskaya2021}. As can be seen in Figure~\ref{fig:T-and-Tu}, the temperature $T$ reaches a maximum in the region of QGP phase and is about 200 MeV for the time of maximal overlap in the central cut.

The characteristic Unruh temperature, as can be seen from the $z = 0\,$fm profiles in Fig.~\ref{fig:T-and-Tu}, has a maximal value at early times after the nuclei touch $\tdiff = 0.5\,$fm$/c$. Then the value of the characteristic Unruh temperature rapidly decreases with the evolution and expansion of the resulting medium. The most accelerated medium in the central cut is formed mainly outside the fireball, where the medium consists of hadrons. In the QGP region, $\Tu$ has a very small value at the initial time after the collision and practically disappears by the time of maximum overlap. Thus, in the central cut the cold regions outside the fireball, consisting of hadrons, are highly accelerated. The received pictures exhibit core-corona structures.

The qualitative picture of the evolution of the temperature $T$ and the characteristic Unruh temperature $\Tu$ remains the same for both central collisions ($b = 0\,$fm, Fig.~\ref{fig:T-and-Tu}(a)) and off-center ones ($b = 7.5\,$fm, Fig.~\ref{fig:T-and-Tu}(b)). The main difference is that for the non-central collisions the medium volume decreases and the acceleration is slightly higher than for the central ones.

\subsection{Comparison: partonic and hadronic phases}

\begin{figure}
	\includegraphics[width=0.9\linewidth]{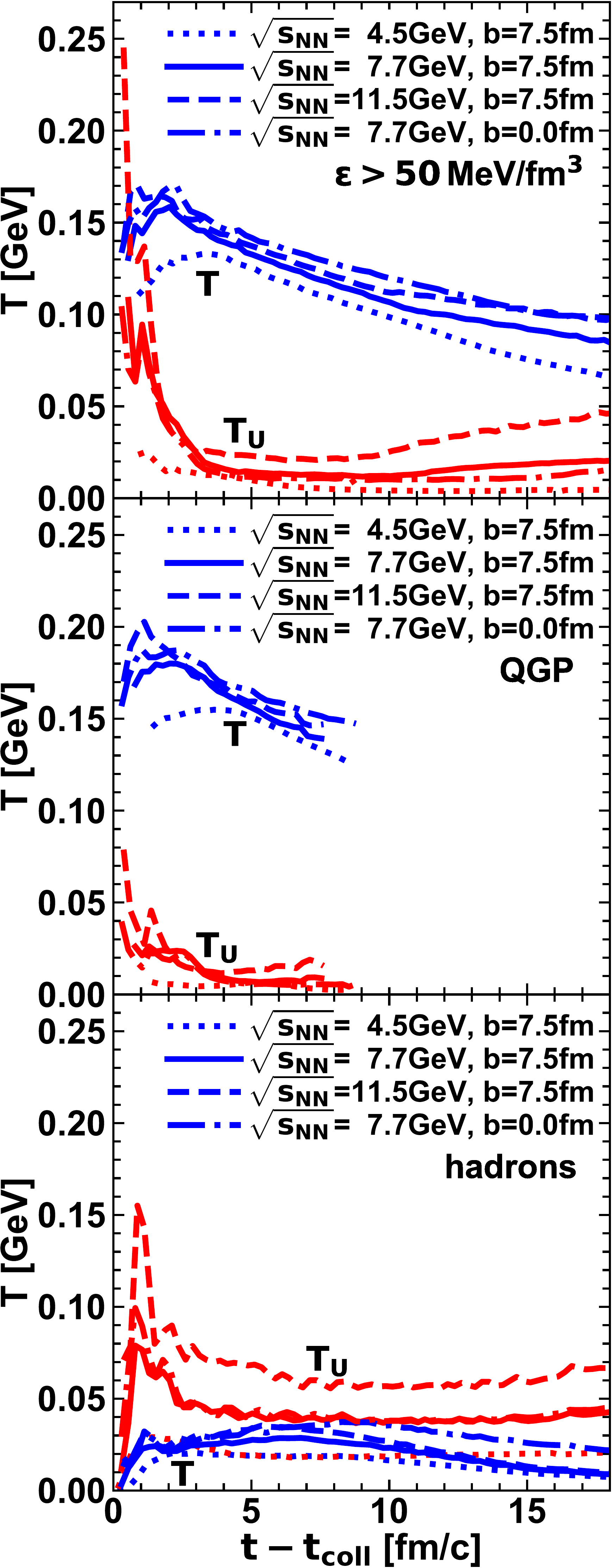}
	\caption{Evolution of the temperature $T$ and the characteristic Unruh temperature $\Tu$, calculated in PHSD for Au + Au collisions at energies $\sqrt{s_{NN}} = 4.5, 7.7$ and $11.5\,$GeV. For the collision energy $\sqrt{s_{NN}} = 7.7\,$GeV, two impact parameters are considered: $b = 0\,$fm, which corresponds to a central collision, and $b = 7.5\,$fm, which corresponds to an off-center collision of nuclei. The results are presented for a full fireball $\varepsilon > 50\,$MeV/fm$^3$ (first panel); for partonic phase only (second panel) and for hadronic phase only (third panel).}
	\label{fig:T-time}
\end{figure}

To better understand the physical picture presented in Figure~\ref{fig:T-and-Tu} at a qualitative level, we calculated a time evolution of the temperatures $T$ and $\Tu$ separately for partonic and hadronic phases, and averaged over the entire volume of the fireball. Such plots for Au + Au collisions at energies $\sqrt{s_{NN}} = 4.5 - 11.5\,$GeV  are presented in Figure~\ref{fig:T-time}.

Let us first discuss the results obtained for the collision energies $\sqrt{s_{NN}} = 7.7$ and $11.5\,$GeV for the fireball (Fig.~\ref{fig:T-time} (first panel)). After the moment of contact of the nuclei, when the fireball is just beginning to form, the characteristic Unruh temperature has a maximum value. Then the value of $\Tu$ decreases rapidly, which is associated with the rapid expansion of the fireball. The temperature $T$ of the fireball has a maximum value at the moment of maximum overlap of the nuclei, which corresponds to $\tdiff \approx 2.5\,$fm$/c$. Then the medium begins to cool, which is associated with the processes of hadronization and thermalization. Note that for the collision energies $\sqrt{s_{NN}} = 7.7$ and $11.5\,$GeV, the obtained maximum value of the temperature $T$ of the resulting medium at the moment of maximum overlap is $T \approx 160 - 170\,$MeV, which is close to the temperature of the confinement-deconfinement phase transition of $150 - 170\,$MeV~\cite{Satz:2018oiz,CastorinaKharzeevSatz2007}.

For the collision energy $\sqrt{s_{NN}} = 7.7\,$GeV, we performed calculations for two values of the impact parameters $b = 0\,$fm and $b = 7.5\,$fm. It can be seen that the values and evolution of the characteristic Unruh temperature do not depend on the centrality of the collision, while the thermodynamic temperature $T$ takes higher values for the central collisions. For the collision energy $\sqrt{s_{NN}} = 
7.7\,$GeV, the characteristic Unruh temperature averaged over all events and over the entire volume of the fireball is less than the thermodynamic temperature $T$ of the resulting medium at all times after the nuclei touch for both the central ($b = 
0\,$fm) and off-center ($b = 7.5\,$fm) collisions.  For $\sqrt{s_{NN}} = 11.5\,$GeV at the moment of nuclei touch in a very short time interval ($\tdiff \approx 0.5\,$fm$/c$) the characteristic Unruh temperature $\Tu$ is greater than the temperature $T$. Note that for $\sqrt{s_{NN}} = 11.5\,$GeV, the maximum value of $\Tu \approx 250\,$MeV exceeds the maximum value of the medium temperature at the moment of maximum nuclei overlap $T \approx 170\,$MeV and corresponds to the acceleration $a \approx 1.6\,$GeV. For a collision energy of $\sqrt{s_{NN}} = 4.5\,$GeV in the resulting medium, the characteristic Unruh temperature is significantly less than the temperature $T$ and after the maximum overlap of nuclei ($\tdiff \approx 5\,$fm$/c$) decreases practically to zero.

In the partonic phase, the temperature $T$ is significantly higher than the Unruh temperature $\Tu$ (Fig.~\ref{fig:T-time} (second panel)), and the latter is very small. In contrast, in the hadronic phase, as can be seen in Fig.~\ref{fig:T-time} (third panel), at the initial moments of time after the collision, the Unruh temperature is significantly higher than $T$. The Unruh temperature rapidly decreases by the time of maximum overlap, but remains higher than $T$ up to times $\tdiff \approx 5\,$fm$/c$. Thus, states with temperature $T$ below the Unruh temperature are found in the hadronic phase, while the medium containing partons is hot and unaccelerated.

\subsection{Analysis of $T^2-T_U^2$}

\begin{figure*}
	\centering
	\includegraphics[width=0.85\linewidth]{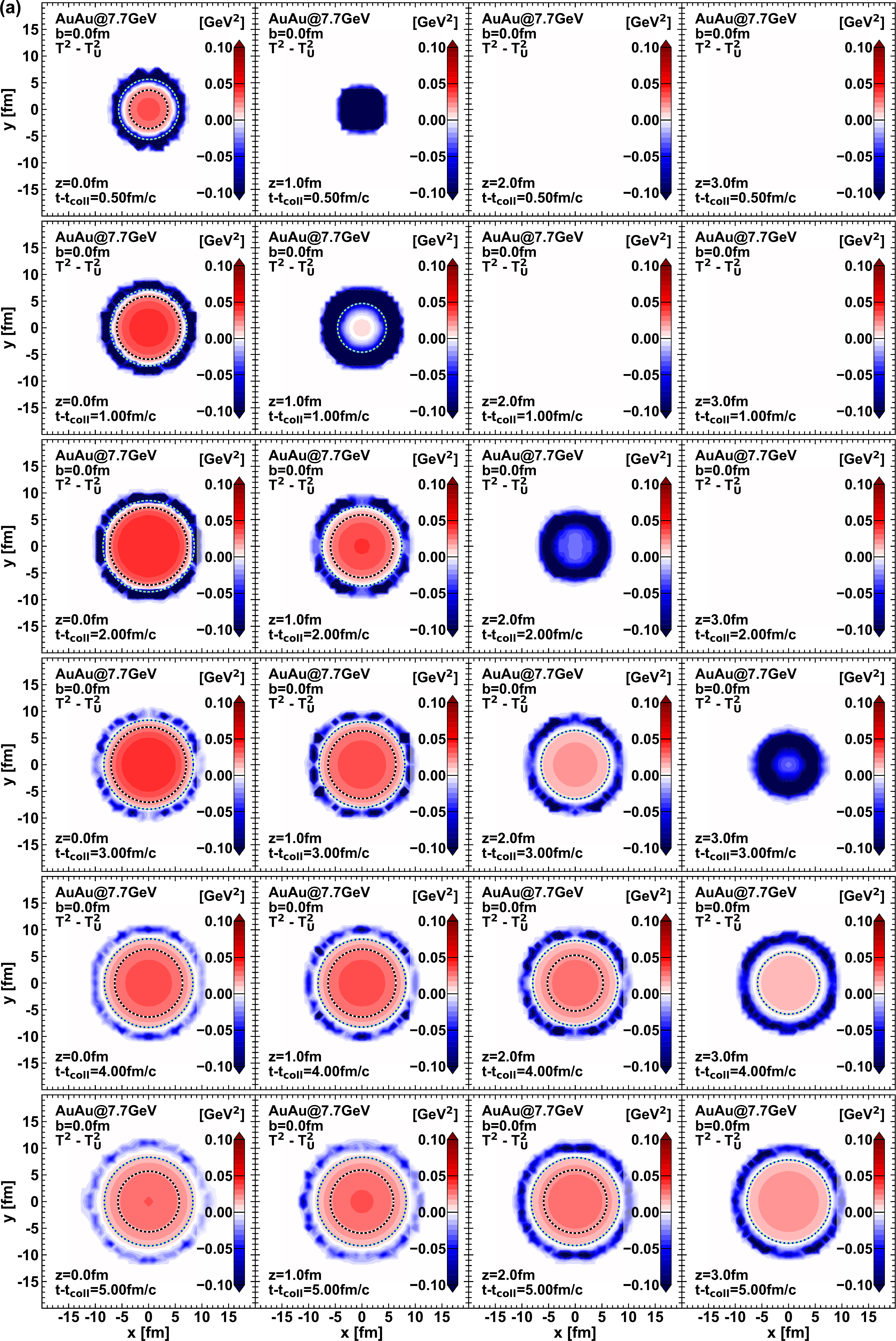}
	\caption{$\Tdiff$ profiles for different $z$-cuts calculated at various times after the collision for Au + Au central collisions ($b=0\,$fm) at $\sqrt{s_{NN}} = 7.7\,$GeV, averaged over all events. Only positive $z$ is shown due to symmetry. The dotted curves on the $\Tdiff$ profiles show two contours in the $xy$ plane, limiting the medium in energy density to the value $\varepsilon = 500\,$MeV/fm$^3$ (inner contour) and $\varepsilon = 50\,$MeV/fm$^3$ (outer contour).}
	\label{fig:T2-Tu2-b0.0}
\end{figure*}

\begin{figure*}
	\centering
	\includegraphics[width=0.85\linewidth]{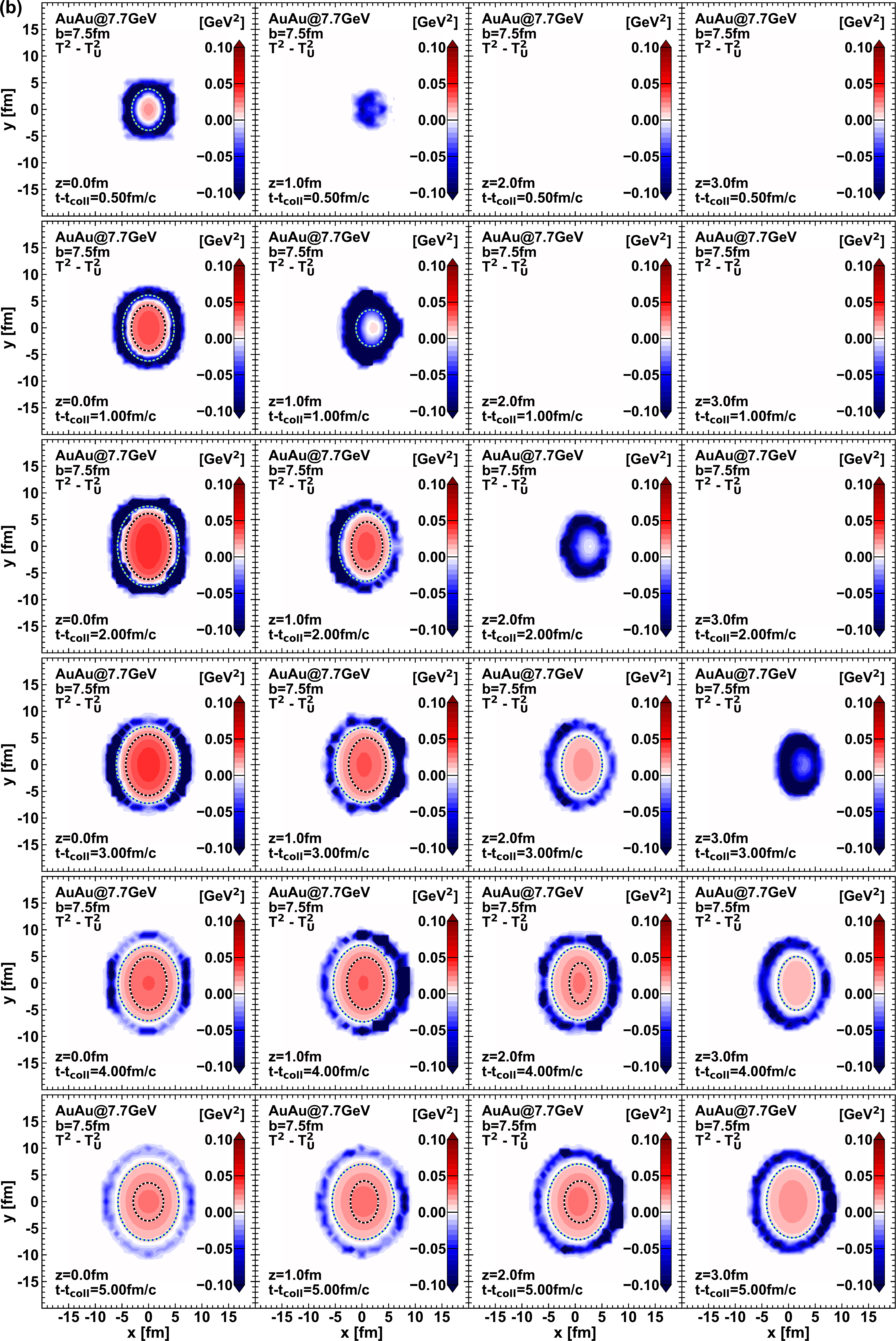}
	\caption{The same as Fig.~\ref{fig:T2-Tu2-b0.0} but for the non-central collisions ($b=7.5\,$fm).}
	\label{fig:T2-Tu2-b7.5}
\end{figure*}

Since we analyze heavy-ion collisions from the point of view of observing and interpreting the phase transition when $T^2 < \Tu^2$, the greatest interest is in the analysis of the difference of the squares of the corresponding temperatures $\Tdiff$. We show the profiles of the difference $\Tdiff$ for various $z$-cuts ($z \geq 0$), calculated at various times after the collision of nuclei for both the central ($b = 0$\,fm, Fig.~\ref{fig:T2-Tu2-b0.0}) and off-center ($b = 7.5$\,fm, Fig.~\ref{fig:T2-Tu2-b7.5}) collisions. Let us first consider the central cut ($z = 0\,$fm). In the central cut the difference $\Tdiff$ is negative mainly in the outside the fireball while inside the fireball $\Tdiff > 0$. There is a very small region of the fireball near the boundary, which disappears by the time $\tdiff = 3 - 4\,$fm$/c$ after the nuclei touch, where $\Tdiff < 0$. In the case of an off-center collision, the region inside the fireball, in which $\Tdiff < 0$, occupies a somewhat larger relative part of the volume, compared to the case $b = 0\,$fm. For an off-center collision at the time $\tdiff = 3\,$fm$/c$ inside the fireball regions are still observed in which $\Tdiff < 0$. In both the central collision and the off-center collision by the time $\tdiff = 5\,$fm$/c$ inside the fireball the difference $\Tdiff$ is positive. Inside the QGP phase region the difference $\Tdiff$ is positive for both the central and off-center collisions.

Let us now discuss the evolution of $\Tdiff$ in the volume of the medium. Due to symmetry, we will consider only cuts with $z > 0\,$fm. It can be seen that in the first moments of time after the collision ($\tdiff = 0 - 1\,$fm$/c$) the $\Tdiff$ profiles depend significantly on $z$. The fireball is formed inside the central cell ($z = 0\,$fm). In the regions with $z > 0\,$fm the medium remains cold and accelerated and does not extend beyond $z = 1\,$fm. The QGP phase is concentrated in the central cell ($z = 0\,$fm) while the regions with $z > 0\,$fm contain hadrons. After $\tdiff = 1\,$fm$/c$ the dependence of the difference $\Tdiff$ on $z$ weakens and starting from $2\,$fm$/c$ after the collision the qualitative picture of the difference $\Tdiff$ does not change depending on $z$. The fireball expands and the outer regions remain accelerated. In the presented profiles it can be seen that the transition from the region with $\Tdiff > 0$ to the region with $\Tdiff < 0$ occurs through the region $\Tdiff = 0$. In the central section the latter corresponds to the transition from the region of the QGP phase to the hadron phase at the moment of maximum overlap. Thus, the hot medium containing the QGP lies inside the cold accelerated region consisting of hadrons. The transition between the phases occurs through the region $\Tdiff = 0$.

\subsection{The direction of acceleration}

\begin{figure*}
	\centering
	\includegraphics[width=0.5\linewidth]{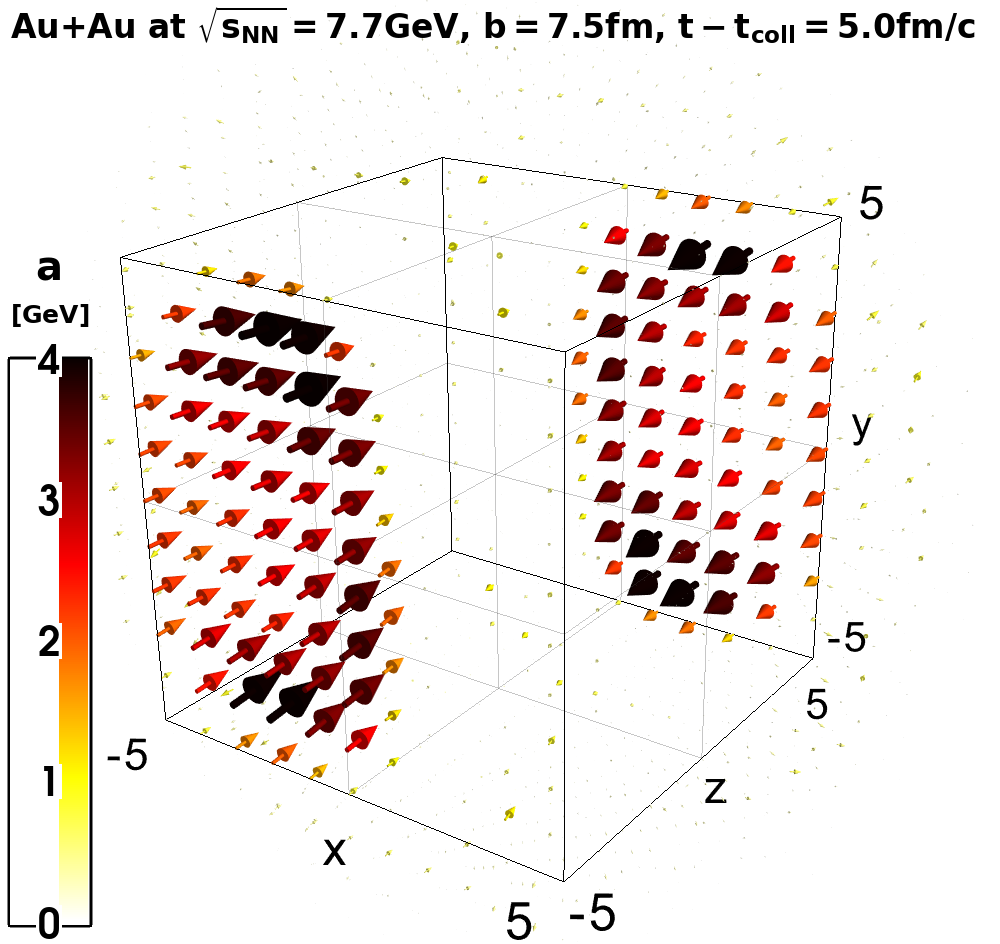}
	\caption{Three-dimensional distribution of the three-dimensional part of the acceleration vector $ a^{\mu} = u^{\nu}\partial_{\nu} u^{\mu} $ for a non-central collision  $ b=7.5\,$fm at $ t-t_{coll} =5\,$fm/$ c $.}
	\label{new1}
\end{figure*}

\begin{figure*}
	\centering
	\includegraphics[width=0.9\linewidth]{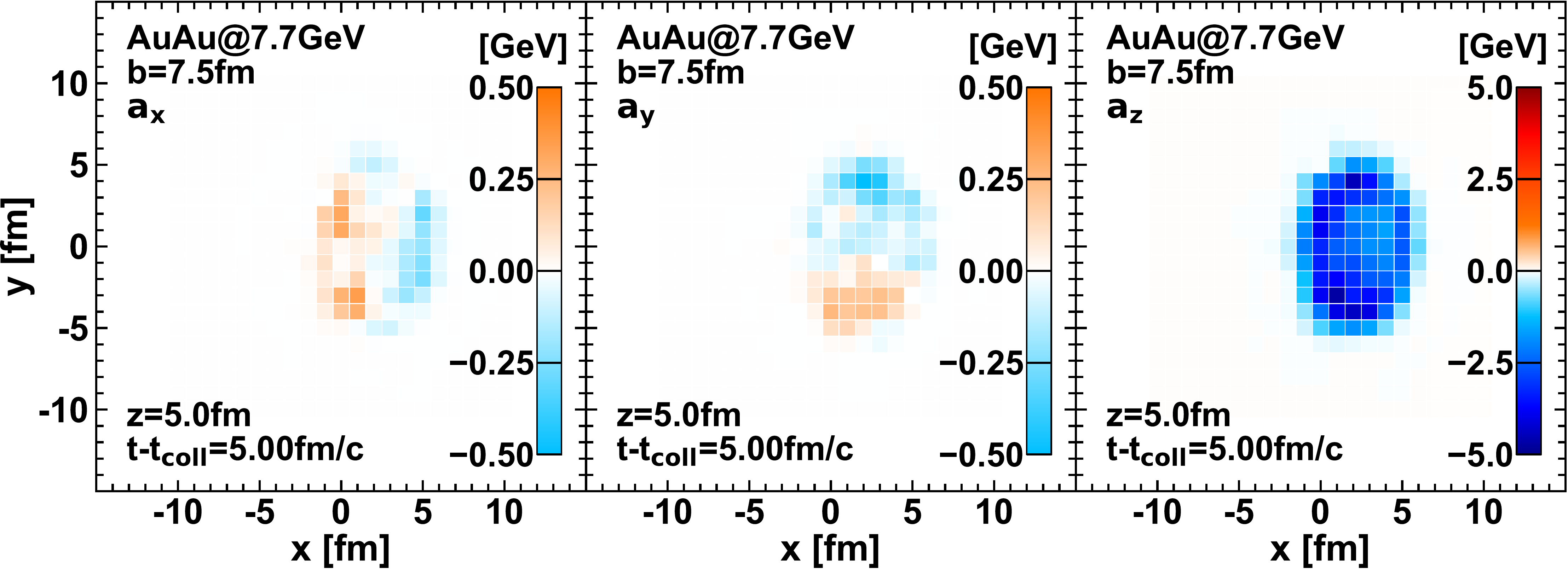}
	\caption{Various projections of the three-dimensional part of acceleration $ a^{\mu} = u^{\nu}\partial_{\nu} u^{\mu} $ for a non-central collision $ b=7.5\,$fm at $ z= 5\,$fm and $ t-t_{coll} =5\,$fm/$ c $.}
	\label{new3}
\end{figure*}

Let's also briefly discuss where the acceleration is directed. Typical distribution for the three-dimensional part of a four-dimensional vector $ a^{\mu} = (a^0, \bold{a}) = (\gamma \dot{\gamma} c, \gamma^2 \frac{d \bold{v}}{d t} + \gamma \dot{\gamma} \bold{v})  $, where $ \gamma = 1/\sqrt{1- \bold{v}^2/c^2} $ is the relativistic gamma factor, is shown in the Figure \ref{new1}. Figure \ref{new3} also shows the distribution of different projections. Let's consider the core and the corona separately.


\begin{itemize}[label={\tiny\textbullet}, leftmargin=*]  
\item \textbf{Central region}. The acceleration of the central dense and hot region is aligned with the direction of the system’s expansion. At small $ z $ the transversal component of the acceleration dominates.
At larger values of $z$ the $ z $-component of the acceleration dominates, increasing to the boundaries of the hot zones. Nonetheless, this acceleration remains at least an order of magnitude lower than that observed in the corona.

\item \textbf{Corona area}. The acceleration in the corona area is mostly opposite directed to the direction of the system’s expansion, especially at values of $ z $ close to the maximums. For such zones the $ z $-component of the acceleration is an order of magnitude larger than the other projections. However, actually, in the areas most distant from the center, the acceleration structure becomes more chaotic and requires deeper analysis.
\end{itemize}

It is important to note that the verbal description presented above is significantly simplified. Evidently, the acceleration field demonstrates significant variations in both space and time. Furthermore, at lower collision energies, the acceleration field will be substantially influenced by the presence of spectators. This will result in the additional deceleration along the boundary between the spectators and participants. We plan to study this issue in more detail in the future.

\section{Discussion}\label{sec:interpretation}

\subsection{Interpretation: phase transition at Unruh temperature}

The obtained results confirm the statements about the properties of accelerated and thermal systems predicted recently at the theoretical level~\cite{Prokhorov:2023dfg}.

The Unruh effect (and the Minkowski vacuum) corresponds to the straight line $T=a/2\pi$ on the phase diagram $(a, T)$. The case of arbitrary $a$ and $T$ can be considered by embedding the system into the Euclidean Rindler space described by the metrics
\begin{align}\label{eq:rindler-metric}
	ds^2 = \frac{\rho^2}{\nu^2} d\varphi^2 + dx^2 + dy^2 + d\rho^2,
\end{align}
where the usual (Euclidean) Rindler coordinates $\rho$ and $0 < \varphi < 2\pi$ are introduced, see e.g.~\cite{Dowker1994, Dowker:1977zj}. The coordinate $\rho = a^{-1}$, that is, it sets the inverse proper acceleration, and the proper temperature is included into the parameter $\nu = 2 \pi T/a$. This space is the direct product of a 2-dimensional cone and a 2-dimensional plane $M = R^2 \otimes C^2_\nu$.

If the acceleration is not too high, then $T > a/2\pi$, and this region has been studied very well in many works, see e.g.~\cite{Dowker1994, FrolovSerebryanyi1987, Morita2019}. And at $T > a/2\pi$, the total angle (measured along the cone surface) turns out to be less than $2\pi$, that is, the angular deficit $\theta_{cone}$ is positive, as well as energy density $\varepsilon$:
\begin{align}
	T > \Tu :\,\,\,\, \theta_{cone} = 2\pi (1 - a/2\pi) > 0, \,\,\,\,  \varepsilon(a,T) > 0.
\end{align}
Also the trace of the stress-energy tensor is zero $T^\mu_\mu = \varepsilon - 3p = 0$ (in the ultrarelativistic limit).  At the same time, the $T < a/2\pi$ region has been studied much less; in particular, one of the few works is~\cite{Prokhorov:2023dfg}. It turns out that this area has very unusual properties. Firstly, the total angle of the cone turns out to be larger than $2\pi$, e.g. the angular deficit becomes negative. Also, the energy density in this case turns out to be negative
\begin{align}\label{eq:theta<0}
	T < \Tu :\,\,\,\, \theta_{cone} < 0,\,\,\,\, \varepsilon(a,T) < 0.
\end{align}
and $T^\mu_\mu \neq 0$ now. Note that this energy is generally not equivalent to the fluid energy, for example, in equations (\ref{eq:Tmunu-sum}) and (\ref{eq:Tmunu-eigen}) and contains the contribution of particles and vacuum, the separation of which is a non-trivial problem. Therefore, the negativeness of the energy density (\ref{eq:theta<0}) does not contradict the fact that in the PHSD model the energy density is positive.

At the point $T = a/2\pi$ the heat capacity undergoes a discontinuity, that is, a second-order phase transition occurs
\begin{align}\label{eq:phase-transition}
	\frac{d\varepsilon}{dT}\Big\vert_{T \rightarrow \Tu + 0} \neq
	\frac{d\varepsilon}{dT}\Big\vert_{T \rightarrow \Tu - 0}.
\end{align}
Such unusual properties of the region $T < a/2\pi$ could cast doubt on how feasible they are in nature. However, the negativity of energy in~\eqref{eq:theta<0} simply means the predominance of ``potential'' energy associated with acceleration over ``kinetic'' energy associated with temperature. As already mentioned, energy depends on both temperature $ T $ and on acceleration $ a $. The first of these contributions, ``kinteic energy'', includes the usual energy of medium at a certain temperature $ T $ (for example, in the case of free massless particles it is given by the Stephen-Boltzmann law). And the second contribution, ``potential energy'', is related with effective statistical interactions with acceleration, described by boost generator (for more details see \cite{Becattini2018, Prokhorov:2023dfg}), and can be negative. We are interested in the situation when the second contribution exceeds the first.

The critical behavior~\eqref{eq:phase-transition} is due to the fact that at the Unruh temperature the lower Matsubara modes become singular at the event horizon $ \rho=0 $. These modes can be obtained as eigenfunctions of the square of the Dirac operator $ \slashed{D}^2 \phi = \lambda \phi $ in the Euclidean Rindler space (\ref{eq:rindler-metric}). It is essential that for an every $ n $-th Matsubara mode there are always two solutions of this equation
\begin{eqnarray} \label{modes}
\phi^{\pm}_{n,s_1} \sim 
e^{i \pi T (2n+
1)\tau} J_{\pm[\frac{2 \pi T}{a} (n+1/2)-s_1/2]}(\xi \rho)\,,
\label{two sol}
\end{eqnarray}
but only one of these solutions, $ \phi^{+} $ or $ \phi^{-} $, is finite at the horizon $ \rho \to 0 $, since Bessel function asymptotically $ J_{b}(x) \to x^{b} $ at $ x\to 0 $. In the formula (\ref{modes}) $ s_1 = \pm 1 $ is the eigenvalue of the boost operator, $ \tau =\rho \varphi/\nu $ is the imaginary proper time in Rindler coordinates and $ \xi $ is defined by momenta and $ \lambda $. 

\begin{figure}
	\centering
	\includegraphics[width=0.9\linewidth]{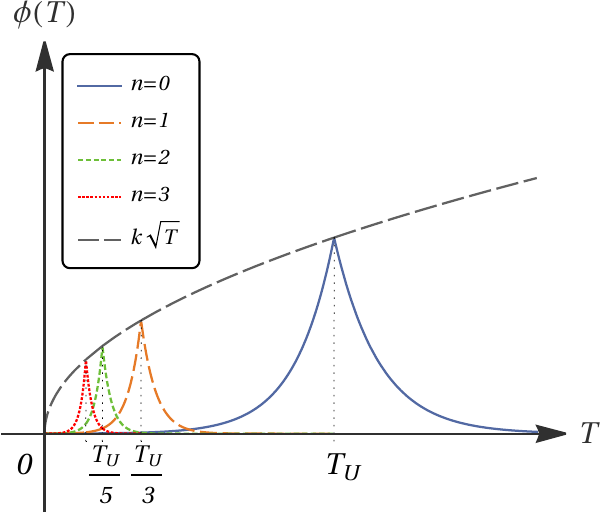}
	\caption{Change in the behavior of $ n $-th Matsubara modes when crossing temperatures $ T=T_U/(2 k +1) $ with $ k=0,1... $. The significant temperature-dependent part of the solution with $ s_1 =1 $ is shown (without exponential factor).}
	\label{new2}
\end{figure}

The key point is that when passing to the region $ T<T_U $, the finite solutions $ \phi^{+}_{0,1} $ and $ \phi^{-}_{-1,-1} $ becomes singular and vice versa. Thus, we change the solution to the modes $ \phi^{-}_{0,1} $ and $ \phi^{+}_{-1,-1} $ , which leads to a jump-like behavior of the Green function and the mean values. This ultimately leads to the phase transition (\ref{eq:phase-transition}). This situation is repeated for temperatures $ T=T_U/(2 k+1) $, $ k=0,1... $ . Note that although the described mode behavior was observed in the simplest case of massless free fields, it can be generalized to the massive case, and it is probably a fairly general phenomenon, like the Unruh effect itself. The jump-like behavior of the modes is shown in the Figure \ref{new2}.

It was suggested \cite{Prokhorov:2023dfg} that such states with $ T<T_U $ are formed in the early stages of the collision of heavy ions, when, due to deceleration of ions, the beams interaction region is characterized by large acceleration, but has not yet had enough time to thermalize. This statement is well confirmed now by the simulation results shown in Figures~\ref{fig:T-and-Tu}-\ref{fig:T2-Tu2-b7.5}: really in the early stages of the collision, the temperature is below the Unruh temperature.

In view of the results obtained, the following question is raised: is it possible to draw conclusions about the existence of a phase transition? Here we also come to an interesting observation. As explicitly shown in Figure~\ref{fig:T-time} the region $T < a/2\pi$, at the same time corresponds predominantly to the hadronic phase, and the region $T > a/2\pi$ corresponds to the partonic phase or quark-gluon plasma\footnote{Note that there is an interesting correspondence: massive particles, hadrons, lead to a non-zero trace of the stress-energy, just like in phase~\eqref{eq:theta<0}}. Thus, we see that there is a close connection between the phase transition~\eqref{eq:phase-transition} and the confinement-deconfinement transition. 

This coincidence is not accidental, see discussion in~\cite{Prokhorov:2023dfg}. Indeed, initially after the collision $T < a/2\pi$ and the energy is negative (see (\ref{eq:theta<0})), being concentrated in the hadronic strings. The negative energy levels can be filled, with an increase in the energy of the system (with increasing temperature to Unruh temperature) and the simultaneous formation of particles with a thermal spectrum~\cite{Morita2019}. Thus, it was suggested that the state $T < a/2\pi$ decays to form a thermal spectrum of hadrons, thus providing rapid thermalization in heavy ion collisions. Since hadronization occurs as a result of the QCD phase transition, two processes (confinement-deconfinement and phase transition~\eqref{eq:phase-transition} at $T = a/2\pi$) occur simultaneously, or, more preferably, transition~\eqref{eq:phase-transition} might provide a dual description of the QCD phase transition. The obtained modeling results confirm this statement at least at a qualitative level. We note a recent work \cite{Chernodub:2025ovo}, where also the relationship between chiral transition and acceleration was studied using the Nambu-Jona-Lasinio model as an example and a conclusion was made about a direct relationship between critical temperature and acceleration. 

We note that, according to, the Figures \ref{fig:T2-Tu2-b0.0} and \ref{fig:T2-Tu2-b7.5} even at $ t = 5\,$fm/$c$ in the central region $ T>T_U $. Of course, over time, the entire QGP should turn into hadrons. Therefore, it can be assumed that the temperature becomes lower than $T_U$ everywhere.

However, we are still considering not so high times $ t\sim 5 \,$fm/$c $, at which within the framework of the model used the quark-gluon plasma can still be present in the system. Therefore, the fact that in the central region $ T>T_U $ does not contradict the above-described picture of phase separation. Moreover, Figure \ref{fig:T-time} confirms the separation of phases with respect to $ T_U $.
However, it should be noted that at the moment we have only some elements of a complementary description of hadronization based on the discussed new phase transition at $ T_U $ and we do not claim, that hadrons always correspond to $ T<T_U $ (and vice versa for quark-gluon plasma).

However, generally speaking, the interpretation discussed in this Subsection does not claim to accurately reproduce all the results of the PHSD model; moreover, such a correspondence would be surprising, given the completely different physical foundations. But at the same time we see that it allows us to shed light on some of the results obtained.

\subsection{Origin of the large acceleration in the corona}

Thus, as a result of modeling we have shown that a very large acceleration occurs, and, predominantly, the greatest acceleration is achieved in the corona. The natural question is -- what is the source of the large acceleration in the corona?

This is a very interesting and non-trivial question. First of all, it should be noted that a similar structure was also observed in the case of vorticity \cite{Baznat:2015eca} (see also \cite{Ayala:2020soy}).
We have several assumptions about possible sources of large acceleration in the corona:
\begin{itemize}[label={\tiny\textbullet}, leftmargin=*]  
\item \textbf{Relativistic effect}. Large acceleration in the corona region can be attributed to the fact that, at the boundaries of
the system, particles forming the fluids approach speeds close to the speed of light. Consequently, these areas exhibit
a significant Lorentz factor, thereby resulting in deceleration caused by a purely relativistic effect.
\item \textbf{Geometry of the system}. Colliding nuclei at the edges of the system are thinner than in the center. Therefore,
given that the system stops, it should be expected that the regions associated with the corona should stop in a shorter
interaction time than the regions forming the core and therefore experience greater acceleration.
\item \textbf{Luttinger relation}. There is another intriguing possibility: in a state of thermodynamic equilibrium, an effective force arises, associated with the temperature gradient $ a \sim \nabla T $,
known as the ``Luttinger force'' \cite{Luttinger:1964zz, Becattini:2016stj}. It is very close to the Tolman-Ehrenfest equilibrium criterion \cite{Tolman:1930ona} in gravitational field, as well as the conditions of global thermodynamic equilibrium \cite{Becattini:2016stj}. Of course, the system in particle collisions is far from global equilibrium, however, it is possible that Luttinger relation works approximately. It would be interesting if this relation were fulfilled in the core or corona.
\end{itemize}
Thus, unfortunately, we do not have a definitive final answer to the question of the sources of large acceleration in the corona. However, clarifying this issue is an interesting task for future research.

\section{Conclusions}\label{sec:conclusions}

Using the PHSD transport model, we studied the generation of acceleration in both center and off-center collisions of gold ions at the intermediate energy of $\sqrt{s_{NN}} = 4.5 - 11.5\,$GeV. In both cases, we show that extremely strong acceleration of the order of 1 GeV is generated in dynamics. This, due to the equivalence principle, allows us to consider experiments with ions as a kind of ``gravitational laboratory''.

The spatial distribution of the acceleration has a core-corona structure, and we find states with temperatures below the Unruh temperature. These states consist mostly of hadrons, while states with a temperature above the Unruh temperature contain a partonic phase. Short after the collision, the acceleration is maximum in the regions outside the fireball (corona), while the regions of the fireball (core) are practically not accelerated and are hot. Inside the fireball, the temperature $T$ is significantly higher than the Unruh temperature $\Tu$. This qualitative picture is preserved for collision energies $\sqrt{s_{NN}} = 4.5 - 11.5\,$GeV. However, the value of the characteristic Unruh temperature becomes significantly higher at high collision energies. After $5\,$fm$/c$, when the medium has thermalized, we practically do not observe accelerated regions in the center of the system.

We also briefly discussed the issue of the direction of acceleration, in particular, in the central region it is predominantly directed in the direction of expansion of the system.

From a theoretical point of view, states with $T < a/2\pi$ have unusual properties. The corresponding Euclidean metric has the form of a cone with an angle greater than $2\pi$. In this case, the transition through $T = a/2\pi$ is accompanied by a phase transition associated with the singularity of the lowest Matsubara modes on the horizon , and for $ T<\Tu  $ the energy density becomes negative. If we assume that the negative energy states are filled with the formation of a positive energy hadronic thermal spectrum, then the Unruh temperature and the corresponding phase transition should roughly coincide with the QCD phase transition. The observation of phase separation relative to the Unruh temperature supports the described view on the thermal hadron production.

\begin{acknowledgments}
The authors are thankful to Alejandro Ayala, Elena Bratkovskaya, Maxim Chernodub and Evgeni Kolomeitsev for valuable discussions and comments. The calculations has been carried out using the ``Govorun'' supercomputer provided by the Laboratory of Information Technologies (JINR). The work of N. S. Tsegelnik also was supported in part by the grants for young scientists and specialists of JINR № 25-301-06. The work of G. Yu. Prokhorov, D. A. Shohonov and V. I. Zakharov was supported by Russian Science Foundation Grant No 24-22-00124.
\end{acknowledgments}



\bibliography{phsd-unruh}

\end{document}